\documentstyle{lamuphys}
\include{psfig}
\makeatletter
\let\chapter\hid@chapter
\makeatother
\def\spose#1{\hbox to 0pt{#1\hss}}

\def\kms{\ifmmode {\rm\,km\,s^{-1}}\else
    ${\rm\,km\,s^{-1}}$\fi}
\def\kmsMpc{\ifmmode {\rm\,km\,s^{-1}\,Mpc^{-1}}\else
    ${\rm\,km\,s^{-1}\,Mpc^{-1}}$\fi}
\def\kpc{{\rm\,kpc}}

\def\msun{\ifmmode {\rm\,M_\odot}\else ${\rm\,M_\odot}$\fi}
\def\Msun{\ifmmode {\rm\,M_\odot}\else ${\rm\,M_\odot}$\fi}
\def\lsun{\ifmmode {\rm\,L_\odot}\else ${\rm\,L_\odot}$\fi}
\def\Lsun{\ifmmode {\rm\,L_\odot}\else ${\rm\,L_\odot}$\fi}
\def\rsun{\ifmmode {\rm\,R_\odot}\else ${\rm\,R_\odot}$\fi}
\def\Rsun{\ifmmode {\rm\,R_\odot}\else ${\rm\,R_\odot}$\fi}

\def\cm{{\rm\,cm}}
\def\cm3{\ifmmode {\rm\,cm^{-3}}\else ${\rm\,cm^{-3}}$\fi}

\def\ergps{\ifmmode {\rm\,erg\,s^{-1}}\else ${\rm\,erg\,s^{-1}}$\fi}
\def\ergpscm2{\ifmmode {\rm\,erg\,s^{-1}\,cm^{-2}}\else
    ${\rm\,erg\,s^{-1}\,cm^{-2}}$\fi}

\def\eg{{\it e.g.}}
\def\deg{\ifmmode {^{\circ}}\else {$^\circ$}\fi}
\def\degr{\ifmmode {^{\circ}}\else {$^\circ$}\fi}
\def\degs{\ifmmode {^{\circ}}\else {$^\circ$}\fi}

\def\etal{{\it et al.~}}

\def\h3Mpc{h^{-3}{\rm Mpc}^3}
\def\Ho{\ifmmode {\rm\,H_\circ}\else ${\rm\,H_\circ}$\fi}
\def\hnot{\ifmmode {\rm\,H_\circ}\else ${\rm\,H_\circ}$\fi}
\def\h0{\ifmmode {\rm\,H_\circ}\else ${\rm\,H_\circ}$\fi}
\def\hnotunit{\ifmmode {\rm\,km\,s^{-1}\,Mpc^{-1}}\else
    ${\rm\,km\,s^{-1}\,Mpc^{-1}}$\fi}
\def\lya{{\rm\,Ly-$\alpha$~}}

\def\qnot{\ifmmode {\rm\,q_\circ}\else ${\rm q_\circ}$\fi}
\def\q0{\ifmmode {\rm\,q_\circ}\else ${\rm q_\circ}$\fi}

\def\arcsec{\ifmmode {^{\prime\prime}~}\else $^{\prime\prime}~$\fi}
\def\asec{\ifmmode {^{\prime\prime}}\else $^{\prime\prime}$\fi}
\def\arcmin{\ifmmode {^{\prime}}\else $^{\prime}$\fi}
\def\amin{\ifmmode {^{\prime}}\else $^{\prime}$\fi}

\def\secper{\ifmmode \rlap.{^{s}}\else $\rlap{.}{^{s}} $\fi}
\def\minper{\ifmmode \rlap.{^{m}}\else $\rlap{.}{^m} $\fi}
\def\magper{\ifmmode \rlap.{^{m}}\else $\rlap{.}{^m} $\fi}
\def\arcsper{\ifmmode \rlap.{^{\prime\prime}}\else
    $\rlap.{^{\prime\prime}}$\fi}
\def\arcmper{\ifmmode \rlap.{^{\prime}}\else
    $\rlap.{^{\prime}}$\fi}
\def\spose#1{\hbox to 0pt{#1\hss}}
\def\simlt{\mathrel{\spose{\lower 3pt\hbox{$\mathchar"218$}}
     \raise 2.0pt\hbox{$\mathchar"13C$}}}
\def\simgt{\mathrel{\spose{\lower 3pt\hbox{$\mathchar"218$}}
     \raise 2.0pt\hbox{$\mathchar"13E$}}}

\def\aa{{A\&A}}
\def\aasup{{A\&AS}}

\def\aj{{AJ}}
\def\apj{{ApJ}}

\def\apjl{{ApJ}}

\def\apjs{{ApJS}}

\def\mnras{{MNRAS}}
\def\nature{{Nature}}

\def\apjref#1;#2;#3;#4 {\par\pp#1, {#2}, #3, #4 \par}

\begin{document}
\pagenumbering{arabic}
\title{Very High Redshift Radio Galaxies}

\author{Wil\,van Breugel\inst{1}, Carlos\,De Breuck\inst{1,2}, 
Huub\,R\"ottgering\inst{2}, George\,Miley\inst{2}, and Adam Stanford\inst{1}}

\institute{University of California Inst. of Geophysics and 
Planetary Physics, LLNL L-413, P.O. Box 808, Livermore, CA 94550, USA
\and
Leiden Observatory, P.O. Box 9513, 2300 RA, Leiden, The Netherlands}

\authorrunning{Wil van Breugel \etal}
\maketitle

\begin{abstract}
High redshift radio galaxies (HzRGs) provide unique targets for the
study of the formation and evolution of massive galaxies and galaxy
clusters at very high redshifts. We discuss how efficient HzRG samples
are selected, the evidence for strong morphological evolution 
at near-infrared wavelengths, and for jet-induced star formation
in the $z = 3.800$ HzRG 4C41.17.
\end{abstract}

\section{Introduction}

Radio sources are surprisingly effective beacons for identifying
galaxies at extremely high redshifts.  Optical/near--IR campaigns
during the past few years by several groups have resulted in the
discovery of more than 120 radio galaxies at $z > 2$, including 17
with $z > 3$, and 3 with $z > 4$.  At low redshifts powerful radio
galaxies are uniquely identified with massive ellipticals. If this is
true also at high redshift, as seems reasonable given the surprisingly
good Hubble $K-z$ relation for radio galaxies at $0 < z < 4.4$, then
we should be able to systematically study the evolution of massive
elliptical galaxies over large lookback times using samples selected
by their radio emission.  While recently developed techniques of
finding very distant star--forming galaxies (\eg\ Steidel \etal\ 1996)
are yielding substantial galaxy populations at $z \sim 3$, radio
galaxy samples remain the best means of finding the most massive
galaxies at $z \sim 3$, and even higher.

Hierarchical galaxy formation scenarios suggest that these massive
galaxies are assembled from smaller structures at relatively late
cosmic epochs (\eg\ Baron and White 1987).  Observations of HzRGs may
thus provide a unique opportunity to study the beginning of this
process.  Furthermore, when they have been found, HzRGs may also be
used for 'color-dropout' searches of galaxy clusters around them, and
begin cluster evolution studies at very high redshift.

\section{How To Find High Redshift Radio Galaxies}

\subsection{Ultra Steep Spectrum Sources}

\begin{figure}[h]
\centerline{
\psfig{figure=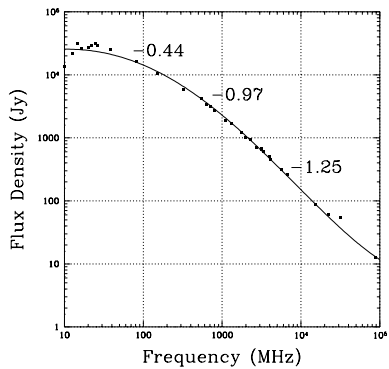,width=5.5cm}
}
\caption{\label{cyga}
\small{The radio spectral energy distribution of Cygnus A, using single dish
measurements from the literature (\eg\ Kellermann, Pauliny-Toth and
Williams 1969, and others). Note the spectral steepening with
frequency.
}}
\end{figure}

\begin{figure}[h]
\centerline{
\psfig{figure=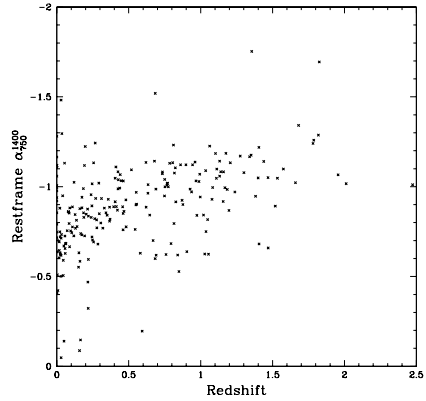,width=5.7cm}
}
\caption{\label{3crrest}
\small{Spectral steepening with redshift for all 3CR radio galaxies.
}}
\end{figure}

It has been known for many years that radio sources with very steep
spectra are mostly associated with faint, distant galaxies (\eg\ Tielens
\etal\ 1979; Blumenthal \& Miley 1979).  An example of a steep radio
spectrum, for the powerful nearby radio galaxy Cygnus A, is shown in
figure \ref{cyga}. One can also see that the radio spectrum steepens at
higher frequency.  If all powerful radio galaxies have such curving
spectra, due to synchrotron and/or inverse Compton losses, then with
increasing redshift we would observe the steeper parts of their
spectra. Analysis of the 3CR radio sample indeed showed such an
expected $\alpha - z$ correlation (van Breugel and McCarthy 1990; Fig
\ref{3crrest} shows an updated version). A similar result has recently
been reported for a sample of ultra-luminous radio galaxies at high
redshifts (Carilli \etal 1998).

Thus this ultra-steep spectrum (USS) `red radio colors' selection
technique may be used as a very effective tool for identifying very
high redshift radio galaxies.  The first comprehensive USS search for
HzRGs was begun by Chambers and collaborators and resulted in the
discovery of the $z = 3.800$ radio galaxy 4C41.17 (Chambers, Miley and
van Breugel 1990). For many years this was the most distant galaxy
known.  This relatively bright object, with its extended \lya halo,
radio-aligned UV continuum, and evidence for star formation (see
below) in many ways stimulated the continued search for more USS
HzRGs.

With the advent of several new, deep radio surveys it is now possible
to define much better USS samples which reach 10 -- 100 times fainter
flux density levels than previous samples, and which allow much more
accurate spectral index determinations. We have used several such
surveys to define USS samples for study in the northern and southern
hemispheres.  Our primary northern hemisphere sample uses the WENSS
325 MHz survey (Rengelink \etal 1997) together with the NVSS 1.4 GHz
survey (Condon \etal 1998), to define the spectral indices, and
the FIRST 1.4 GHz survey (Becker \etal 1995), to obtain radio maps.
In the southern hemisphere we use the low frequency Texas 365 MHz
(Douglas \etal 1996; Dec $> -35$\deg) and MRC 408 MHz (Large \etal 1981;
Dec $< -35$\deg) surveys, in combination with the NVSS and PMN 5 GHz
(Griffith \& Wright 1993; Dec $< -35$\deg) surveys.

The sources in our southern hemisphere sample all have spectral
indices $\alpha < -1.2$.  We have obtained radio images of all our
southern $USS$ targets with the Australia Telescope and VLA so that we
can make accurate identifications. A deep near-infrared identification
program of this sample was begun with the CTIO 4m telescope.
Spectroscopic observations of some objects from this sample with the
ESO 3.6 m have shown that USS sample is indeed extremely efficient at
finding HzRGs, and have already resulted in the discovery of the most
distant radio galaxy in the Southern Hemisphere known to date (TN
J1338$-$1942 at $z=4.13$; De Breuck \etal, these proceedings).  A
summary of the current status of HzRG identification programs is given
in Table 1.

\begin{table}
\begin{center}
\begin{tabular}{| c || c || c | c | c || c |}\hline
\small
Sample & Definition & \multicolumn{3}{c||}{Known Redshift} & Unknown \\ 
\hline
 & & $z<2$ & $2<z<3$ & $z>3$ &  \\
\hline
3CR & S$_{178} >$ 10 Jy & 99.5 \% & 0.5 \% & 0 \% & 1\% \\
MRC/1Jy & S$_{408} >$ 0.95 Jy & 93.4 \% & 5.9 \% & 0.7 \% & 41 \% \\
4C USS & $\alpha_{\small 178}^{\small 1414} < -1.0$ & 53 \% & 35 \% & 12 \% & 50 \% \\
New USS & $\alpha_{\small 325}^{\small 1400} < -1.3$ & 35 \% & 35 \% & 30 \% & 45\% (faint) \\
\hline
\end{tabular}
\end{center}
\caption{
\small{The HzRG content for four samples of radio sources as a function of z: 3CR (Spinrad, private communication), MRC/1Jy (McCarthy \etal\ 1996), 4C (Chambers \etal\ 1996), and our new USS sample.}}
\end{table}
\normalsize

\begin{figure}[h]
\centerline{
\psfig{figure=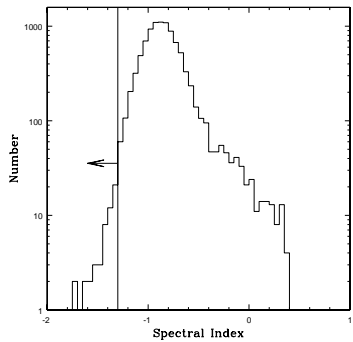,width=4.8cm}
}
\caption{\label{mphis}
\small{Hystogram of southern USS sources at DEC $<$ -35\deg 
from the MRC and PMN surveys.
In total 76 out of 2915 sources have $\alpha^{5 GHz}_{408 MHz} < -1.20$.
}}
\end{figure}

\subsection{The $K-z$ Diagram}

\begin{figure}[h]
\centerline{
\psfig{figure=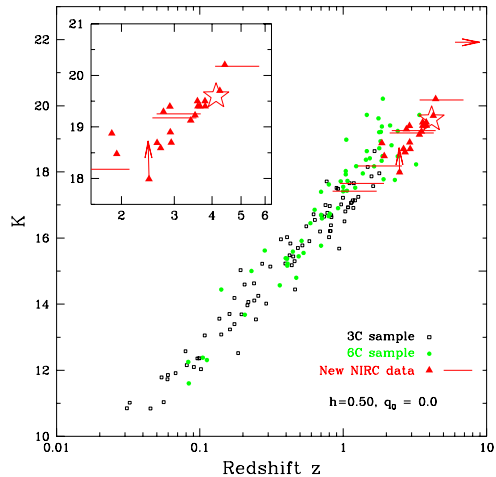,width=7.5cm}
}
\caption{\label{kzdiag} \small{Hubble K-z diagram for the 3C and 6C
surveys (Eales \etal 1997), and NIRC data for HzRGs from van Breugel
\etal\ (1998). The new USS $K$-band detections and very high redshift
candidates are shown by horizontal bars at the predicted redshift. The
horizontal arrow indicates the range of predicted redshift for our
faintest $K$-band object. The open star represents TN J1338-1942 (De
Breuck \etal, this volume). All magnitudes are corrected to a 64 kpc
metric diameter, assuming H$_{0}$ = 50, q$_0$ = 0.  The inset shows a
blowup of the NIRC data in the $1.5 < z < 5.5$ and $17.5 < K < 21$
region.}}
\end{figure}

Around the time that 4C41.17 was discovered another systematic search
for HzRGs also proved to be successful.  This method uses the
magnitude -- redshift relationship at infrared wavelengths, the $K-z$
`Hubble' diagram, for powerful radio galaxies galaxies to find
promising targets. This resulted in the discovery of the $z = 3.395$
radio galaxy B2 0902+34 (Lilly 1988). Indeed, the $K-z$ diagram has
since proved to be a very reliable, if little understood,
tool. Despite strong morphological evolution seen at near-infrared
wavelengths, discussed below, the most powerful radio sources continue
to follow the $K-z$ relationship even to z $\sim$ 4.4 (Fig
\ref{kzdiag}).  It is our experience that, by combining the USS and
$K-z$ techniques and using near-infrared identifications of USS
sources which are unidentified on UKST/POSS2 plates, one is virtually
guaranteed to be successful in HzRG hunts.  Without near-IR
identifications and photometry approximately 2/3 of the galaxies are
at $z > 2$. With near-IR identifications and photometry one can select
the redshift range one wishes to study and to date all galaxies
observed at Keck were found to be at the redshift predicted from the
$K-z$ diagram, at least for $z \simlt 4.4$.

\section{Morphological evolution of HzRGs}

\begin{figure}[h]
\centerline{
\psfig{figure=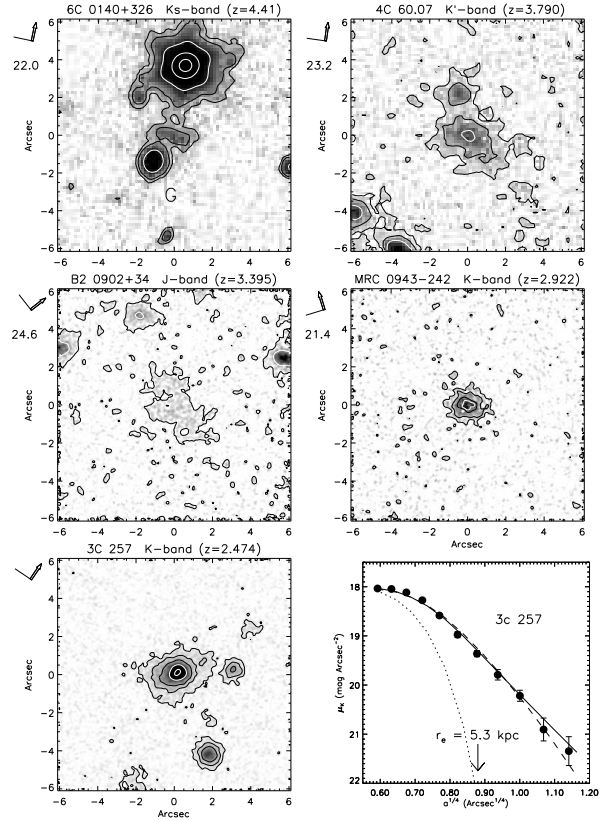,width=7.8cm}
}
\caption{\label{morphev}
\small{Selected near--IR images of HzRGs, presented in order of decreasing
redshift, and the surface brightness profile of 3C257.
}}
\end{figure}

Near-infrared images obtained with the W.\ M.\ Keck I Telescope of
HzRGs with $1.9 < z < 4.4$ show strong morphological evolution at {\it
rest--frame optical} ($\lambda_{\rm rest} > 4000$ \AA) wavelengths (van
Breugel \etal 1998; Fig \ref{morphev}).  At the highest redshifts, $z
> 3$, the rest--frame visual morphologies exhibit structure on at
least two different scales: relatively bright, compact components with
typical sizes of $\sim$10~\kpc\ surrounded by large--scale ($\sim$ 50
$-$ 100 \kpc) diffuse emission. The brightest components are often
aligned with the radio sources, and their {\it individual}
luminosities are $M_B \sim -20$ to $-22.$ For comparison,
present--epoch L$_\star$ galaxies and, perhaps more appropriately,
ultra-luminous infrared starburst galaxies, have, on average, $M_B \sim
-21.0$. The {\it total, integrated} rest--frame B--band luminosities
are $3 - 5$ magnitudes more luminous than present epoch $L_\star$
galaxies.

At lower redshifts, $z < 3$, the rest--frame optical morphologies
become smaller, more centrally concentrated, and less aligned with the
radio structure.  Galaxy surface brightness profiles for the $z < 3$
HzRGs are much steeper than those of at $z > 3$.  We attempted to fit
the $z < 3$ surface brightness profiles with a de Vaucouleurs
r$^{1/4}$ law and with an exponential law, the forms commonly used to
fit elliptical and spiral galaxy profiles, respectively.  We
demonstrate the fitting for our best resolved object at $z < 3$, 3C
257 at $z = 2.474$ (Fig \ref{morphev}).  Within the limited dynamical
range of the data, both functional forms fit the observed
profiles---neither is preferred.  Interestingly, despite this strong
morphological evolution the $K - z$ `Hubble' diagram for the most
luminous radio galaxies remains valid even at the highest redshifts,
where a large fraction of the K-band continuum is due to a
radio--aligned component.

\section{Jet-induced star formation in 4C41.17: HST imaging}

\begin{figure}[ht]
\centerline{
\psfig{figure=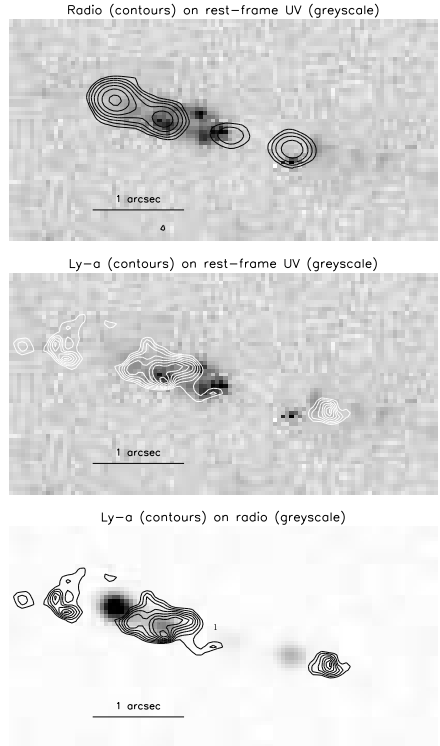,width=7.7cm}
}
\caption{\label{4c4117hst}
HST WFPC2 images of 4C41.17-North with the X-band map of Carilli \etal 
1994 superimposed.
}
\end{figure}

\begin{figure}[ht]
\centerline{
\psfig{figure=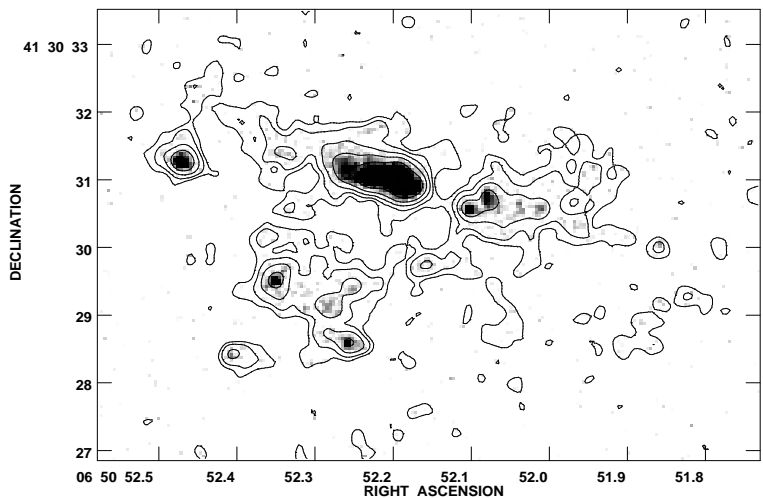,width=10.5cm}
}
\caption{\label{4c4117smooth}
Smoothed version of the HST WFPC2/F702W image 
showing the entire 4C41.17 system, including the clumpy companion system 
4C41.17-South.
}
\end{figure}

Deep HST images of 4C41.17 at `R'-band (rest-frame UV) and in \lya
show that the 4C41.17 system consists of two components: 4C41.17-North
with a bright string of UV knots and \lya emission along the radio
axis, and 4C41.17-South with several much fainter UV knots,
distributed in random fashion throughout a low surface brightness
halo. The brightest radio knot in 4C41.17-North is associated with the
brightest UV knot and arc-like \lya emission. One of the field objects
in the HST images was also seen at near--IR (Graham \etal 1994; object
\# 16) and radio wavelengths (Carilli \etal 1994). This was used to
align the HST and radio frames with an estimated relative accuracy of
$\sim$ 0.1\arcsec.  The central, radio--aligned UV and \lya emission
is shown in Figure \ref{4c4117hst} with the 0.21\arcsec resolution
radio X-band image from Carilli \etal overlaid. Figure
\ref{4c4117smooth} shows the HST continuum image of the entire 4C41.17
system smoothed to 0.3\arcsec resolution to enhance low surface
brightness features.

Downstream from the bright radio/UV/\lya knot the radio source curves
towards a faint, very steep spectrum NE lobe (see Carilli \etal 1994),
while upstream from this knot, towards the central radio core, the UV
continuum appears edge--brightened. These morphological features
suggest a strong interaction between the radio jet and dense ambient
gas and, in fact, are as expected in jet--induced star formation
models where sideways shocks induce star formation in the dense medium
of forming galaxies (\eg\ De Young 1989; Bicknell \etal 1998).

The star formation rates ($SFR$) of the various components as
deduced from the rest--frame UV HST photometry range from 5 -- 40
\msun/yr for the knots in 4C41.17-North, to 5 -- 10 \msun/yr in
4C41.17-South.  Here we have assumed L$_{1500\AA} \sim 10^{40.1}$
\ergps\AA$^{-1}$ for a $SFR$ = 1 \msun/yr (Conti \etal 1996) and no
dust reddening. The derived values are surely lower limits, given the
detection of dust at sub---mm wavelengths in 4C41.17 by Dunlop \etal
(1994).

The entire 4C41.17 system is embedded in a common halo of diffuse, low
surface brightness emission which extends over a very large area of
54$h_{50}^{-1}$~kpc $\times$ 76$h_{50}^{-1}$~kpc (5\arcsec $\times$
7\arcsec). This includes a faint region, 4C41.17-South, with half a
dozen compact knots distributed in random fashion. Spectroscopic
observations have shown that 4C41.17-South is indeed at the same
redshift as 4C41.17-North (Dey \etal 1999).  The range of UV
luminosities and $SFR$ rates for the individual knots in
4C41.17-South is lower than in 4C41.17-North, and very similar to the
`normal' (radio-quiet) Lyman-break galaxies discovered by Steidel \etal
(1996). The random distribution and on average lower $SFR$ in the
4C41.17-South knots suggests that star formation here is unaided by
bowshocks from the radio jet.  The total star formation rate, integrated
over the entire 4C41.17 system and including the low surface brightness
emission, is $\sim$ 660 \msun/yr.  Of this perhaps as much as 2/3 of
the star formation may be occurring in the inter-knot regions. If the
total star formation would continue at this rate for $2 \times 10^{8} -
2 \times 10^{9}$ yrs an entire massive elliptical galaxy of $10^{11}
\msun - 10^{12} \msun$ might be assembled between $z \sim 4$ and $z
\sim 2.5$, consistent with the morphological evolution for HzRGs seen
in the near--IR Keck observations.

\section{Jet-Induced Star Formation In 4C41.17: Keck Spectroscopy}

\begin{figure}[ht]
\centerline{
\psfig{figure=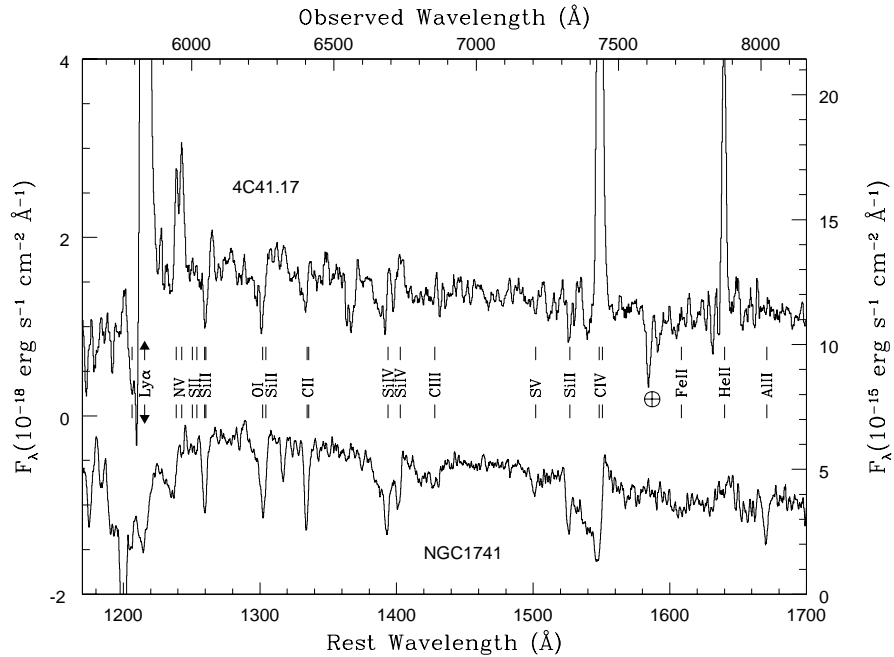,width=13cm}
}
\caption{\label{4c4117spec}
Keck spectrum (Dey \etal 1997) of the radio-aligned component
4C41.17-North, compared with a UV spectrum of the Wolf-Rayet starburst
galaxy NGC 1741 (Conti \etal 1996).
}
\end{figure}

Deep spectropolarimetric observations with the Keck II telescope by
Dey \etal (1997) have provided strong evidence in support of the
jet--induced star formation model for 4C41.17-North suggested above on
the basis of the HST and radio morphologies. These observations showed
that the bright, radio--aligned rest--frame UV continuum is unpolarized
($P_{UV}(2\sigma) < 4\%$).  This implies that scattered AGN light,
which is generally the dominant contributor to the rest-frame UV
emission in $z\sim 1$ radio galaxies, is unlikely to be a major
component of the UV flux from 4C 41.17. Instead, the total light
spectrum shows absorption lines and P--Cygni--like features that are
similar to those detected in the spectra of the recently discovered
population of star forming galaxies at slightly lower ($z\sim2-3$)
redshifts (Fig \ref{4c4117spec}).  The detection of the S V$\lambda 1502$ stellar
photospheric absorption line, the shape of the blue wing of the Si IV
profile, the unpolarized continuum emission, the inability of other
AGN--related processes to account for the UV continuum flux, and the
overall similarity of the UV continuum spectra of 4C 41.17 and the
nearby star forming region NGC 1741B strongly suggest that the light
from 4C 41.17 is dominated by young, hot stars.  The presence of
radio--aligned features in many of the $z > 3$ HzRGs suggests, by
analogy to 4C41.17, that jet--induced star formation may be a common
phenomenon at these very high redshifts.

\noindent {\bf Acknowledgments} 

{\small Part of the work described here was performed while WvB was on
sabbatical leave during January 1997 -- April 1997 at the
Anglo-Australian Observatory, the Australian National Telescope
Facility, and the Mount Stromlo and Sidings Springs Observatories. He
appreciates the support provided by these institutes. He thanks the
Kookaburra's for their wake--up calls, and his Australian colleagues
for their warm hospitality and invigorating discussions, with special
thanks to Drs.\ J.\ Bland--Hawthorn, G.\ Bicknell, M.\ Dopita, and R.\
Sutherland.  We gratefully acknowledge S.\ Rawlings for advance
information regarding 6C~0140+326 and 8C~1435+635 and his work on the
$K - z$ diagram, and C.\ Carilli for providing high quality radio
images of 4C41.17 which allowed to improve on the relative
radio/optical astrometry for this source. The research by WvB, CDB and
AS at IGPP/LLNL is performed under the auspices of the US Department
of Energy under contract W--7405--ENG--48.}

%
%

\end{document}